\documentclass[twocolumn,showpacs,preprintnumbers,amsmath,amssymb]{revtex4}
\usepackage{graphicx}

\begin{document}

\title{New type of rogue waves}
\author{N.\,V. Ustinov}
\email{n_ustinov@mail.ru}
\affiliation{Joint Institute for Nuclear Research, 141980 Dubna, Russia}

\begin{abstract}
New type of localized solutions for the two-dimensional multicomponent Yajima--Oikawa system  is presented. 
The dynamics of solutions of this type occurs on the zero background and is similar to that of rogue waves. 
\end{abstract}
\maketitle

\begin{center}
\bf\small I.\quad INTRODUCTION 
\end{center}

Much attention of researchers has been paid in the recent decades to the study of rogue  waves \cite{OOS,DKM,AAT,KPS,ORBMA,DDEG,BCGWC,ORB,CBSGM,DGMCD,SPP,LSM}. 
Various mechanisms of formation of these waves were suggested. 
The occurrence of rogue waves is most often investigated on the basis of the mechanisms of modulation instability and superposition of waves \cite{OOS,ORBMA,BCGWC,ORB,SPP,LSM}. 
In both cases, an evolution of rogue waves takes place against the background of a wave field, which is reflected in the definitions of such waves \cite{DKM,ORBMA,ORB}. 
In this report, the localized waves developed in the absence of the background wave fields are considered. 
At the same time, their dynamics corresponds to the dynamics of rogue waves that "appear from nowhere and disappear without a trace" \cite{AAT}. 

A search among solutions of the multi-dimensional nonlinear equations for ones suitable for describing the behavior of rogue waves is of great interest. 
The solutions having dynamics similar to the dynamics of rogue waves were obtained as particular cases of lump (rational) solutions, semi-rational ones and their generalizations (see, e.g., Refs.~\cite{DM,MS,OY1,OY2,LQMH}). 
It is important to find other types of solutions describing the dynamics of rogue waves. 
The mechanisms generating such waves may be different. 

The investigation of the two-dimensional multicomponent Yajima-Oikawa (YO) system attracts significant attention in the recent years 
\cite{OMO,RKLG,KVSL,SK,KVL,KKT,CCFM_1,CCFM_2,CCFMa,RPHK,SU4}. 
This system comprises multiple (say $N$) short-wave components and a single long-wave one. 
It generalizes the scalar ($N=1$) two-dimensional YO system \cite{ZMNP} and is often called the 2D coupled long-wave--short-wave resonance interaction system.  

The two-dimensional multicomponent YO system belongs to the class of equations integrable by the inverse scattering transformation method \cite{M}. 
Also, it arose in different physical contexts. 
The two-component system and the multicomponent one were derived by applying the reductive perturbation method in Refs.~\cite{OMO} and \cite{KVL}, respectively, as the governing equations for the interaction of dispersive waves in a weak Kerr-type nonlinear medium. 
In these systems, the short waves propagate in anomalous dispersion regime while the long wave propagates in the normal dispersion regime. 
A generation of the terahertz radiation by optical pulses in a medium of asymmetric quantum particles is described under the quasi-resonance conditions by the two-dimensional two-component YO system \cite{SU4}.
 
Various types of solutions of the two-dimensional multicomponent YO system were found. 
So, rational and semi-rational solutions mentioned above due to their role under considering rogue waves were investigated in Refs.~\cite{CCFM_2} and \cite{RPHK}, respectively. 
The rational solutions include the fundamental (simplest) and general (multi- and higher-order) lumps and line rogue waves derived from the lumps under the certain parameter conditions \cite{CCFM_2}. 
It was shown that the fundamental lumps and rogue waves have three different patterns: bright, intermediate and dark states. 
The fundamental semi-rational solutions considered in \cite{RPHK} can describe the fission of a dark soliton into a lump and a dark soliton or the fusion of one lump and one dark soliton into a dark soliton. 
The nonfundamental semi-rational solutions were shown to fall into three subclasses: higher-order, multi- and mixed-type semi-rational solutions.

The solutions discussed above of the two-dimensional multicomponent YO system were found using the bilinear method. 
In this report, we exploit the Darboux transformation (DT) technique \cite{MaSa,GHZ} to obtain the solutions of this system. 
Note that the DT technique was applied to the multicomponent YO systems in the one-dimensional case in Refs.~\cite{SU2,SU3,C,CGSC,LG}. 

The paper is organized as follows. 
The two-dimensional multicomponent YO system of the general form and the corresponding overdetermined system of linear equations are given in Section~2. 
Also, the DT formulas for these systems are presented here.  
New type of localized solutions of the two-dimensional multicomponent YO system on the zero background is considered in Section~3, and the stability of solutions of this type is discussed. 
Concluding remarks are given in Section~4.

\mbox{} 

\begin{center}
\bf\small II.\quad OVERDETERMINED LINEAR SYSTEM AND DARBOUX TRANSFORMATION 
\end{center}

The two-dimensional multicomponent YO system is written in the dimensionless form as  
\begin{equation}
\begin{array}{c}
{\displaystyle\frac{\partial\varphi_n}{\partial t}+\frac{\partial\varphi_n}{\partial y}=i\frac{\partial^2\varphi_n}{\partial x^2}+iu\varphi_n\ \ (n=1,\dots,N),}_{\mathstrut}\\ 
{\displaystyle\frac{\partial u}{\partial t}=\frac{\partial}{\partial x}
\sum\limits_{n=1}^N\sigma_n|\varphi_n|^2}^{\mathstrut},
\end{array}
\label{N_YO}
\end{equation}
where $\varphi_n=\varphi_n(x,y,t)$ and $u=u(x,y,t)$ are the $n$th short-wave and long-wave components, respectively, $\sigma_n=\pm1$ ($n=1,\dots,N$).
In the case of the YO system of the general form, parameters $\sigma_n$ have different signs. 

The two-dimensional multicomponent YO system (\ref{N_YO}) has infinitely many integrals of motion. 
The first few integrals are 
\begin{equation}
\begin{array}{c}
{\displaystyle\int\!\!\!\int\!u\,dx\,dy,\quad 
\int\!\!\!\int\!|\varphi_n|^2\,dx\,dy\ \ (n=1,\dots,N),}_{\mathstrut}\\ 
{\displaystyle\int\!\!\!\int\!\left(u^2+i\sum\limits_{n=1}^N\sigma_n\left[\varphi_n
\frac{\partial\varphi_n^*}{\partial x}-\varphi_n^*\frac{\partial\varphi_n}{\partial x}\right]\right)dx\,dy.}^{\mathstrut}
\end{array}
\label{im}
\end{equation}
Also, Eqs.~(\ref{N_YO}) are represented as the compatibility condition of the overdetermined system of linear equations 
\begin{equation}
\begin{array}{c}
{\displaystyle\frac{\partial^2\psi_1}{\partial x^2}=-i\left(
\frac{\partial\psi_1}{\partial t}+\frac{\partial\psi_1}{\partial y}\right)-
u\psi_1,}_{\mathstrut}\\
{\displaystyle\frac{\partial\psi_{n+1}}{\partial x}=
\frac{\sigma_n}{2}\varphi_n^*\psi_1\ \ (n=1,\dots,N),}^{\mathstrut}
\end{array}
\label{le1}
\end{equation}
and 
\begin{equation}
\begin{array}{c}
{\displaystyle\frac{\partial\psi_1}{\partial t}=-\sum\limits_{n=1}^N\varphi_n\psi_{n+1}
,}_{\mathstrut}\\
{\displaystyle\frac{\partial\psi_{n+1}}{\partial t}+\frac{\partial\psi_{n+1}}{\partial y}=
\frac{i\sigma_n}{2}\left(\varphi_n^*\frac{\partial\psi_1}{\partial x}-
\frac{\partial\varphi_n^*}{\partial x}\psi_1\right)}^{\mathstrut}_{\mathstrut}\\
{(n=1,\dots,N).}^{\mathstrut}
\end{array}
\label{le2}
\end{equation}
Here $\psi_k=\psi_k(x,y,t)$ ($k=1,\dots,N+1$) is the $k$th component of the solution of 
Eqs.~(\ref{le1}) and (\ref{le2}).

Let $\chi_k=\chi_k(x,y,t)$ ($k=1,\dots,N+1$) be the $k$th component of a solution of the overdetermined system (\ref{le1}), (\ref{le2}). 
Then, the differential 1-form 
\begin{equation}
d\,\delta(\chi,\psi)=\delta_x(\chi,\psi)dx+\delta_t(\chi,\psi)dt+\delta_y(\chi,\psi)dy,
\label{ddelta}
\end{equation}
where 
$$
\delta_x(\chi,\psi)=\chi_1^*\psi_1,\quad
\delta_t(\chi,\psi)=-2\sum\limits_{n=1}^N\sigma_n\chi_{n+1}^*\psi_{n+1}, 
$$
$$
\delta_y(\chi,\psi)=i\left(\chi_1^*\frac{\partial\psi_1}{\partial x}-
\frac{\partial\chi_1^*}{\partial x}\psi_1\right)-\delta_t(\chi,\psi), 
$$
is closed; i.e., for a contour $\Gamma$ connecting the points $(x_0,y_0,t_0)$ and 
$(x,y,t)$, integral 
\begin{equation}
\delta(\chi,\psi)=\int\limits_{\Gamma}d\,\delta(\chi,\psi)+C
\label{delta}
\end{equation}
($C$ is a constant) depends only on initial and final points and is independent of a specific choice of contour $\Gamma$. 

The overdetermined system of linear equations (\ref{le1}), (\ref{le2}) is covariant with respect to the DT $\psi_k\to\psi_k[1]$ ($k=1,\dots,N+1$), $\varphi_n\to\varphi_n[1]$ ($n=1,\dots,N$), $u\to u[1]$, where the transformed quantities are defined in the following manner \cite{SU4}:
\begin{equation}
\psi_k[1]=\psi_k-\frac{\delta(\chi,\psi)}{\delta(\chi,\chi)}\chi_k\ \ (k=1,\dots,N+1),  
\label{DT_psi}
\end{equation}
\begin{equation}
\varphi_n[1]=\varphi_n-2\sigma_n\frac{\chi_{n+1}^*\chi_1}{\delta(\chi,\chi)}\ \ (n=1,\,\dots,\,N),
\label{DT_phi}
\end{equation}
\begin{equation}
u[1]=u+2\frac{\partial^2}{\partial x^2}\log\delta(\chi,\chi). 
\label{DT_u}
\end{equation}
Relations (\ref{DT_phi}) and (\ref{DT_u}) define new solution of the system (\ref{N_YO}), while expressions (\ref{DT_psi}) give the components of corresponding solution of the overdetermined system (\ref{le1}), (\ref{le2}). 

\mbox{}

\begin{center}
\bf\small III.\quad ROGUE WAVE TYPE SOLUTIONS 
\end{center}

Let us assume that the initial solution of the YO system (\ref{N_YO}) is the zero background:
$$
\varphi_1=\dots=\varphi_N=u=0.
$$
In this case, we have 
\begin{equation}
\chi_{n+1}=f_n(t-y)\ \ (n=1,\dots,N),
\label{chi_0_j}
\end{equation}
where $f_n(t-y)$ ($n=1,\dots,N$) are arbitrary functions of their argument. 
The complex variants of the source function of the heat equation can be used to express the component $\chi_1$ of solution of the overdetermined system (\ref{le1}), (\ref{le2}). 
In the simplest case, this component is written as 
\begin{equation}
\chi_1=\frac{1}{\sqrt{y-\mu}}\,\,\exp\!\left(\frac{i(x-\lambda)^2}{4(y-\mu)}\right),
\label{chi_0_1_sf}
\end{equation}
where $\lambda$ and $\mu$ are complex constants. 
Then, using Eqs.~(\ref{delta}), (\ref{ddelta}), (\ref{chi_0_j}) and (\ref{chi_0_1_sf}), we obtain 
\begin{equation}
\begin{array}{c}
{\displaystyle\delta=\delta(\chi,\chi)=
\sqrt{\frac{\pi}{2\mu_I}}\,\exp\!\left(\frac{\lambda_I^2}{2\mu_I}\right)}_{\mathstrut}\\
{\displaystyle\mbox{}\times\mbox{\rm erf}\!
\left(\frac{\lambda_I(y-\mu_R)-\mu_I(x-\lambda_R)}{\sqrt{2\mu_I}\,|y-\mu|}
\right)}_{\mathstrut}^{\mathstrut}\\
{\displaystyle\mbox{}+2\!\!\int\limits_{t_0-y_0}^{t-y}\!\sum\limits_{n=1}^{N}
\sigma_n|f_n(\zeta)|^2d\zeta+C_0,}^{\mathstrut}
\end{array}
\label{delta_0_sf}
\end{equation}
where $\lambda_R=\Re(\lambda)$, $\lambda_I=\Im(\lambda)$, $\mu_R=\Re(\mu)$, 
$\mu_I=\Im(\mu)>0$, $C_0$ is a real constant, $\mbox{\rm erf}\,(\zeta)$ is the error function. 

After substitution of the expressions (\ref{chi_0_j})--(\ref{delta_0_sf}) into the DT formulas (\ref{DT_phi}), (\ref{DT_u}), we find the following solution of the 
two-dimensional multicomponent YO system (\ref{N_YO}):
\begin{equation}
\varphi_n=-2\sigma_n
\frac{f_n(t-y)^*\,{\rm e}^{\frac{i(x-\lambda)^2}{4(y-\mu)}}}
{\sqrt{y-\mu}\,\delta}\ \ (n=1,\,\dots,\,N),
\label{phi_0_sf}
\end{equation}
\begin{equation}
u=2\frac{\partial^2}{\partial x^2}\log\delta\,.
\label{u_0_sf}
\end{equation} 
It is supposed in what follows that the functions $f_n(t-y)$ and constant $C_0$ are such that the solution (\ref{phi_0_sf}), (\ref{u_0_sf}) is nonsingular. 

Different types of solutions of the two-dimensional YO system (\ref{N_YO}) are obtained by choosing the functions $f_n(t-y)$ ($n=1,\dots,N$) in 
Eqs.~(\ref{delta_0_sf})--(\ref{u_0_sf}) in different manner. 
If, for example, $f_n(t-y)\sim\exp[\varepsilon(t-y)]$ ($\varepsilon$ is a constant) or $f_n(t-y)\to0$ at $|t-y|\to\infty$ ($n=1,\dots,N$) then solution 
(\ref{phi_0_sf}), (\ref{u_0_sf}) is localized on the $(x,y)$-plane for any $t$ and $\varphi_n\to0$ at $|t|\to\infty$. 

Consider an interesting case when parameters $\sigma_n$ ($n=1,\,\dots,\,N$) have different signs and $|f_n(t-y)|\to\infty$ at $|t-y|\to\infty$ ($n=1,\dots,N$). 
Let us assume for the sake of concreteness that 
\begin{equation}
f_n(t-y)=\alpha_n{\rm e}^{\varepsilon_1(t-y)}+\beta_n{\rm e}^{\varepsilon_2(t-y)}\ \ (n=1,\,\dots,\,N),
\label{f_n_exp}
\end{equation}
where $\alpha_n$, $\beta_n$ ($n=1,\dots,N$), $\varepsilon_1$ and $\varepsilon_2$ are complex constants. 
If $\Re(\varepsilon_1)\Re(\varepsilon_2)<0$, then the solution of YO system (\ref{N_YO}), which is obtained after the substitution of expressions 
(\ref{f_n_exp}) into Eqs.~(\ref{delta_0_sf})--(\ref{u_0_sf}), is localized on the $(x,y)$-plane, and, what is particularly important, $\varphi_n\to0$ 
($n=1,\dots,N$) and $u\to0$ at $|t|\to\infty$. 
So, we have localized solution having zero temporal asymptotics. 
Such kind of the dynamics resembles that of rogue waves. 

It is supposed here that the YO system  (\ref{N_YO}) is of general form. 
In the opposite case, when all parameters $\sigma_n$ ($n=1,\,\dots,\,N$) have the same sign, using expressions (\ref{f_n_exp}) leads to the singular solution of the YO system. 

To illustrate the dynamics of the solutions discussed above we consider the simplest case $N=2$, $\sigma_1=1$ and $\sigma_2=-1$.
Eqs.~(\ref{delta_0_sf})--(\ref{f_n_exp}) give us the following expressions for the solution of the two-component YO system: 
\begin{equation}
\varphi_n=-2\sigma_n
\frac{\alpha_n^*{\rm e}^{\varepsilon_1^*(t-y)}+\beta_n^*{\rm e}^{\varepsilon_2^*(t-y)}}{\sqrt{y-\mu}\,\Delta}\,{\rm e}^{\frac{i(x-\lambda)^2}{4(y-\mu)}}\ \ 
(n=1,\,2),
\label{phi_0_2}
\end{equation}
\begin{equation}
u=2\frac{\partial^2}{\partial x^2}\log\Delta\,,
\label{u_0_sf_2}
\end{equation} 
where 
$$
\begin{array}{c}
{\displaystyle\Delta=
\sqrt{\frac{\pi}{2\mu_I}}\,{\rm e}^{\frac{\lambda_I^2}{2\mu_I}}\mbox{\rm erf}\!
\left(\frac{\lambda_I(y-\mu_R)-\mu_I(x-\lambda_R)}{\sqrt{2\mu_I}\,|y-\mu|}
\right)}_{\mathstrut}\\
{\displaystyle\mbox{}+2\!\!\int\limits_{t_0-y_0}^{t-y}\!\sum\limits_{n=1}^{2}
\sigma_n\left|\alpha_n{\rm e}^{\varepsilon_1\zeta}+
\beta_n{\rm e}^{\varepsilon_2\zeta}\right|^2d\zeta+C_0.}^{\mathstrut}
\end{array}
$$

The profiles of the absolute value of component $\varphi_1$ and component $u$ of solution 
(\ref{phi_0_2}), (\ref{u_0_sf_2}) for different values of variable $t$ and for the parameter values $\lambda=i$, $\mu=2i$, $y_0=t_0=0$, $\alpha_1=1$, $\beta_1=2$, $\alpha_2=2$, $\beta_2=1$, $\varepsilon_1=-1$, $\varepsilon_2=1$ and $C_0=6$ are presented in Figs.~1 and 2. 
\begin{figure}
\centering
\includegraphics[angle=0,width=2.7in]{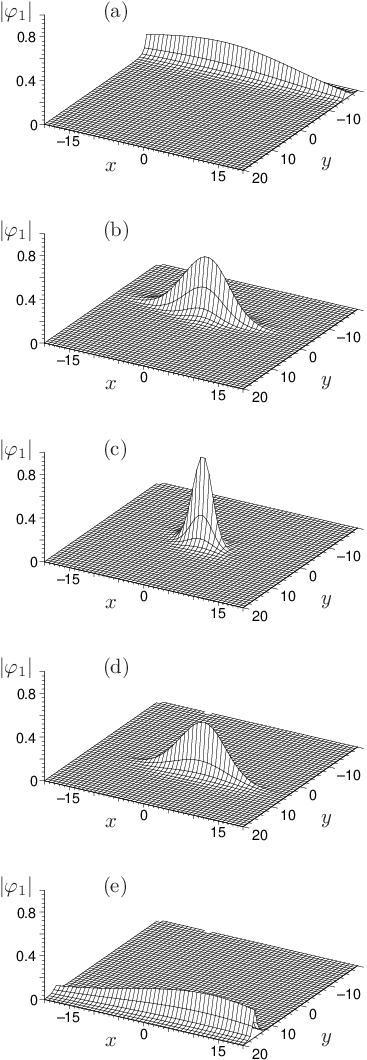}
\caption{Profiles of $|\varphi_1|$ for $t=-16$ (a), $t=-4$ (b), $t=0$ (c), $t=4$ (d) and $t=16$ (e).}
\end{figure}
The complete dynamics is given in the files SM1.gif and SM2.gif in Supplemental Material \cite{SM}. 
It is seen that this solution has form of the solitary wave, and all its components are localized on the $(x,y)$-plane for any $t$. 
In the limit $|t|\to\infty$, the amplitudes of components $\varphi_1$ and $\varphi_2$ tend to zero as $1/\sqrt{|t|}$ (see Fig.~1).
The length $l_y$ of the wave along axis $y$ can be estimated as 
$l_y\sim|\Re(\varepsilon_1)|^{-1}+|\Re(\varepsilon_2)|^{-1}$.
For $|t|\gg\sqrt{|\mu|^2+|\lambda|^2}$, the length $l_x$ along axis $x$ exceeds $l_y$ and can be estimated as 
$$
l_x\sim2|t|\left(\sqrt{\lambda_I^2+4\mu_I}-|\lambda_I|\right)/\mu_I.
$$
The decrease of long-wave component $u$ as $|t|\to\infty$ occurs faster than the short-wave ones (see Fig.~2). 
\begin{figure}
\centering
\includegraphics[angle=0,width=2.7in]{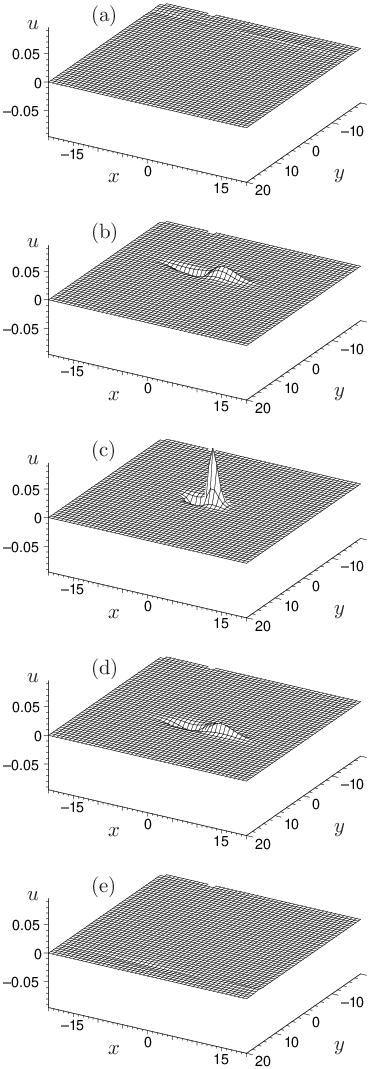}
\caption{Profiles of $u$ for $t=-16$ (a), $t=-4$ (b), $t=0$ (c), $t=4$ (d) 
and $t=16$ (e).} 
\end{figure}

Thus, we see that the dynamics of solitary wave (\ref{phi_0_2}), (\ref{u_0_sf_2}) matches with that of rogue waves \cite{AAT}. 
There is, however, an important distinction. 
Whereas the phenomenon of rogue wave develops on the background waves, the solitary wave (\ref{phi_0_2}), (\ref{u_0_sf_2}) evolves on the zero background. 

The height of rogue wave has to be more than about twice the significant height of  background waves~\cite{DKM,ORBMA,ORB}. 
The waves, whose height exceeds the background value more than five times, are sometimes called super rogue waves \cite{CHOA,SPSCHOA}.
Here the background waves are absent. 
The maximum values of amplitudes of $\varphi_1$, $\varphi_2$ and $u$ of the solitary wave (\ref{phi_0_2}), (\ref{u_0_sf_2}) depend on its parameters and can be arbitrary large. 

Solitary waves having similar dynamics exist for arbitrary number $N>1$ of the short-wave components of system (\ref{N_YO}). 
The functions $f_n(t-y)$ ($n=1,\dots,N$) in Eqs.~(\ref{delta_0_sf})--(\ref{u_0_sf}) have to satisfy conditions $|f_n(t-y)|\to\infty$ at $t-y\to\pm\infty$ in this case. 
For example, these functions can be chosen in accordance with Eqs.~(\ref{f_n_exp}). 
Different signs among $\sigma_n$ ($n=1,\,\dots,\,N$) are necessary to obtain the nonsingular solutions in that case. 

The generalizations of rogue wave of the form (\ref{phi_0_2}), (\ref{u_0_sf_2}) can be obtained if some generalizations of the complex variant of the source function (\ref{chi_0_1_sf}) are used as component $\chi_1$ in the DT formulas. 
In particular, component $\chi_1$ can be chosen in the following manner: 
\begin{equation}
\chi_1=\sum\limits_{m=1}^{M}\sum\limits_{l=1}^{L}\frac{\nu_{lm}}{\sqrt{y-\mu_m}}
\exp\left(\frac{i(x-\lambda_l)^2}{4(y-\mu_m)}\right),
\label{chi_0_g1}
\end{equation}
where $\nu_{lm}$, $\lambda_l$ and $\mu_m$ ($l=1,\,\dots,\,L$; $m=1,\,\dots,\,M$) are complex constants, $\Im(\mu_m)>0$. 
Also, we can put 
\begin{equation}
\chi_1=\left(c_1\frac{\partial}{\partial\lambda}+c_2\frac{\partial}{\partial\mu}\right)
\frac{1}{\sqrt{y-\mu}}\exp\left(\frac{i(x-\lambda)^2}{4(y-\mu)}\right), 
\label{chi_0_g2}
\end{equation}
where $c_1$ and $c_2$ are constants. 
The study of such generalizations of rogue wave  (\ref{phi_0_2}), (\ref{u_0_sf_2}) (multi- and higher-order waves) and their interaction with waves of other types requires a separate consideration.

Note that the stability of rogue wave (\ref{phi_0_2}), (\ref{u_0_sf_2}) with respect to the perturbations of a special kind can be established within the frameworks of the DT technique. 
Indeed, let us take the solution of the overdetermined system (\ref{le1}), (\ref{le2}) in the form 
\begin{equation}
\chi_1=\frac{1}{\sqrt{y-\mu}}\,\,\exp\!\left(\frac{i(x-\lambda)^2}{4(y-\mu)}\right)+\kappa\tilde\chi_1,
\label{chi_0_1_sf_k}
\end{equation}
\begin{equation}
\chi_{n+1}=\alpha_n{\rm e}^{\varepsilon_1(t-y)}+\beta_n{\rm e}^{\varepsilon_2(t-y)}+\kappa F_n(t-y)\ \ (n=1,\,2),
\label{chi_n_exp_k}
\end{equation}
where $\kappa$ is parameter considered to be small, $\tilde\chi_1$ is defined as $\chi_1$ in Eq.~(\ref{chi_0_g1}) or in Eq.~(\ref{chi_0_g2}), $F_{1,2}(t-y)$ are the functions of their argument such that $|F_{1,2}(t-y)|<1$. 
The substitution of expressions (\ref{chi_0_1_sf_k}),  (\ref{chi_n_exp_k}) into the DT formulas (\ref{DT_phi}), (\ref{DT_u}) gives us the perturbed solution of the two-dimensional YO system (\ref{N_YO}). 
This solution coincides with rogue wave (\ref{phi_0_2}), (\ref{u_0_sf_2}) in the case $\kappa=0$. 
It is important that the difference between the perturbed solution and rogue wave 
(\ref{phi_0_2}), (\ref{u_0_sf_2}) will be insignificant during the time evolution if $|\kappa|\ll1$. 
This indicates the stability of rogue wave considered with respect to the perturbations of special form. 

The existence of integrals of motion (\ref{im}) is important in the investigation of stability of rogue wave (\ref{phi_0_2}), (\ref{u_0_sf_2}) in the general case and in the numerical simulations. 
Also, the integrals of motion can be helpful under the study of blowing up of solution (\ref{phi_0_sf}), (\ref{u_0_sf}) that takes place for some values of its parameters. 

\begin{center}
\bf\small IV.\quad CONCLUSION 
\end{center}

In this paper, the new type of rogue waves for the two-dimensional multicomponent 
Yajima--Oikawa system is presented. 
The waves of this type are distinguished by the fact that their dynamics occur on the zero background. 
This implies that rogue waves presented here are formed solely due to the nonlinear focusing. 
It seems very important to extend this type of rogue waves to other models of various physical contexts describing the wave interactions.


\begin{thebibliography}{99}
\bibitem{OOS} M. Onorato, A.R. Osborne, M. Serio, Modulational instability in crossing sea states: a possible mechanism for the formation of freak waves, Phys. Rev. Lett. 96 (2006) 014503.
\bibitem{DKM} K. Dysthe, H.E. Krogstad, P. Muller, Oceanic rogue waves, Annu. Rev. Fluid Mech. 40 (2008) 287--310.
\bibitem{AAT} N. Akhmediev, A. Ankiewicz, M. Taki, Waves that appear from nowhere and disappear without a trace, Phys. Lett. A 373 (2009) 675--678.
\bibitem{KPS} C. Kharif, E. Pelinovsky, A. Slunyaev, {\it Rogue Waves in the Ocean} (Springer, New York, 2009). 
\bibitem{ORBMA} M. Onorato, S. Residori, U. Bortolozzo, A. Montina, F.T. Arecchi, Rogue waves and their generating mechanisms in different physical contexts, Phys. Rep. 528 (2013) 47--89. 
\bibitem{DDEG} J.M. Dudley, F. Dias, M. Erkintalo, G. Genty, Instabilities, breathers and rogue waves in optics, Nat. Photonics 8 (2014) 755--764. 
\bibitem{BCGWC} F. Baronio, S. Chen, P. Grelu, S. Wabnitz, M. Conforti, Baseband modulation instability as the origin of rogue waves, Phys. Rev. A 91 (2015) 033804.
\bibitem{ORB} M. Onorato, S. Resitori, F. Baronio (Eds.), {\it Rogue and Shock Waves in Nonlinear Dispersive Media}, Lect. Notes Phys. 926 (Springer, Switzerland, 2016).
\bibitem{CBSGM} S.H. Chen, F. Baronio, J.M. Soto-Crespo, P. Grelu, D. Mihalache, Versatile rogue waves in scalar, vector, and multidimensional nonlinear systems, J. Phys. A: Math. Theor. 50 (2017) 463001. 
\bibitem{DGMCD} J.M. Dudley, G. Genty, A. Mussot, A. Chabchoub, F. Diaset, Rogue waves and analogies in optics and oceanography, Nat. Rev. Phys. 1 (2019) 675--689. 
\bibitem{SPP} A.V. Slunyaev, D.E. Pelinovsky, E.N. Pelinovsky, Rogue waves in the sea: observations, physics, and mathematics, Phys. Usp. 66  (2023) 148--172. 
\bibitem{LSM} L. Liu, W.R. Sun, B.A. Malomed, Formation of rogue waves and modulational instability with zero-wavenumber gain in multicomponent systems with coherent coupling, Phys. Rev. Lett. 131 (2023) 093801.
\bibitem{DM} P. Dubard, V.B. Matveev, Multi-rogue waves solutions: from the NLS to the KP-I equation, Nonlinearity 26 (2013) 93--125.
\bibitem{MS} V.B. Matveev, A.O. Smirnov, AKNS and NLS hierarchies, MRW solutions, ${\rm P}_n$ breathers, and beyond, J. Math. Phys. 59 (2018) 091419. 
\bibitem{OY1} Y. Ohta, J.K. Yang, Rogue waves in the Davey--Stewartson equation, Phys. Rev. E 86 (2012) 036604. 
\bibitem{OY2} Y. Ohta, J.K. Yang, Dynamics of rogue waves in the Davey--Stewartson II equation, J. Phys. A: Math. Theor. 46 (2012) 105202.
\bibitem{LQMH} Y. Liu, C. Qian, D. Mihalache, J. He, Rogue waves and hybrid solutions of the Davey-Stewartson I equation, Nonlinear Dyn. 95 (2019)  839857. 
\bibitem{OMO} Y. Ohta, K. Maruno, M. Oikawa, Two-component analogue of two-dimensional long-wave--short-wave resonance interaction equations: a derivation and solutions, J. Phys. A: Math. Theor. 40 (2007) 7659--7672.
\bibitem{RKLG} R. Radha, C.S. Kumar, M. Lakshmanan, C.R. Gilson, The collision of multimode dromions and a firewall in the two-component long-wave-short-wave resonance interaction equation, J. Phys. A: Math. Theor. 42 (2009) 102002. 
\bibitem{KVSL} T. Kanna, M. Vijayajayanthi, K. Sakkaravarthi, M. Lakshmanan, Higher dimensional bright solitons and their collisions in a multicomponent long wave--short wave system, J. Phys. A: Math. Theor. 42 (2009) 115103.
\bibitem{SK} K. Sakkaravarthi, T. Kanna, Dynamics of bright soliton bound states in (2+1)-dimensional multicomponent long wave-short wave system, Eur. Phys. J. Spec. Top. 222 (2013) 641--653.
\bibitem{KVL} T. Kanna, M. Vijayajayanthi, M. Lakshmanan, Mixed solitons in (2+1) dimensional multicomponent long-wave--short-wave system, Phys. Rev. E 90 (2014) 042901. 
\bibitem{KKT} A. Khare, T. Kanna, K. Tamilselvan, Elliptic waves in two-component long-wave--short-wave resonance interaction system in one and two dimensions, Phys. Lett. A 378 (2014) 3093--3101. 
\bibitem{CCFM_1} J.C. Chen, Y. Chen, B.F. Feng, K.I. Maruno, Multi-dark soliton solutions of the two-dimensional multi-component Yajima--Oikawa systems, J. Phys. Soc. Jpn. 84 (2015) 034002. 
\bibitem{CCFM_2} J. Chen, Y. Chen, B.F. Feng, K.I. Maruno, Rational solutions to two- and one-dimensional multicomponent Yajima--Oikawa systems, Phys. Lett. A 379 (2015) 1510--1519.
\bibitem{CCFMa} J.C. Chen, Y. Chen, B.F. Feng, Z.Y. Ma, General bright-dark soliton solution to (2 + 1)-dimensional multi-component long-wave-short-wave resonance interaction system, Nonlinear Dyn. 88 (2017) 1273--1288.
\bibitem{RPHK} J. Rao, K. Porsezian, J. He, T. Kanna, Dynamics of lumps and dark-dark solitons in the multi-component long-wave--short-wave resonance interaction system, Proc. R. Soc. A: Math. Phys. Eng. Sci. 474 (2018) 20170627.
\bibitem{SU4} S.V. Sazonov, N.V. Ustinov, Two-dimensional dynamics of solitons under the conditions of Zakharov--Benney resonance, Bulletin of RAS: Phys. 82 (2018) 1359--1362.
\bibitem{ZMNP} V.E. Zakharov, S.V. Manakov, S.P. Novikov, L.P. Pitaevskii, {\it Theory of Solitons: The Inverse Scattering Method} (Consultants Bureau, New York, 1984).
\bibitem{M} V.K. Melnikov, On equations for wave interactions, Lett. Math. Phys. 7 (1983) 129--136.
\bibitem{MaSa} V.B. Matveev, M.A. Salle, {\it Darboux Transformations and Solitons}  (Springer--Verlag, Berlin--Heidelberg, 1991). 
\bibitem{GHZ} C. Gu, A. Hu, Z. Zhou, {\it Darboux Transformations in Integrable Systems} (Springer, Dordrecht, 2005). 
\bibitem{SU2} S.V. Sazonov, N.V. Ustinov, Propagation of vector solitons in a quasi-resonant medium with Stark deformation of quantum states, JETP 115 (2012) 741--758. 
\bibitem{SU3} S.V. Sazonov, N.V. Ustinov, Vector acoustic solitons from the coupling of long and short waves in a paramagnetic crystal, Theor. Math. Phys. 178 (2014) 202--222. 
\bibitem{C} S.H. Chen, Darboux transformation and dark rogue wave states arising from two-wave resonance interaction, Phys. Let.t A 378 (2014) 1095--1098.
\bibitem{CGSC} S.H. Chen, P. Grelu, J.M. Soto-Crespo, Dark- and bright-rogue-wave solutions for media with long-wave--short-wave resonance, Phys. Rev. E 89 (2014) 011201. 
\bibitem{LG} R. Li, X. Geng, A matrix Yajima--Oikawa long-wave--short-wave resonance equation, Darboux transformations and rogue wave solutions, Commun. Nonlinear Sci. Numer. Simul. 90 (2020) 105408.
\bibitem{SM} See Supplemental Material for the animations SM1.gif and SM2.gif of $|\varphi_1|$ and $u$, respectively, for the parameters of Figs.~1, 2 and $-22\le t\le22$. 
\bibitem{CHOA} A. Chabchoub, N. Hoffmann, M. Onorato, N.~Akh\-mediev, Super rogue waves: observation of a higher-order breather in water waves, Phys. Rev. X 2 (2012)  011015. 
\bibitem{SPSCHOA} A. Slunyaev, E. Pelinovsky, A. Sergeeva, A. Chabchoub, N.~Hoffmann, M.~Onorato, N.~Akhmediev,  Super rogue waves in simulations based on weakly nonlinear and fully nonlinear hydrodynamic equations, Phys. Rev. E 88 (2013) 012909.
\end{thebibliography}
\end{document}